\documentclass[11pt, a4paper]{article}
\pdfoutput=1

\usepackage{amsmath}
\usepackage{amsfonts}
\usepackage{amssymb}
\usepackage{bm}
\usepackage{bbm}
\usepackage{verbatim}
\usepackage{dsfont}
\usepackage{booktabs}
\usepackage{slashed}
\usepackage{nicefrac}
\usepackage{mathtools}
\usepackage{physics}
\usepackage{array}

\usepackage{epsfig}
\usepackage{xcolor}
\usepackage{graphicx}
\usepackage[font=small]{caption}
\usepackage[listofformat=empty,subrefformat=empty]{subfig}

\usepackage{float}
\usepackage{overpic}
\usepackage{tikz}
\usepackage{tabularx}

\usepackage{enumerate}
\usepackage{hhline}
\usepackage{multirow}

\usepackage[nosort]{cite}
\usepackage{xspace}
\usepackage{setspace}

\usepackage[pdftitle={},
  pdfauthor={},
  pdfsubject={},
  bookmarksopen, bookmarksnumbered, bookmarksopenlevel=2, colorlinks=false, linkcolor=blue, citecolor=blue, urlcolor=blue]{hyperref}

\usepackage[nameinlink,capitalize]{cleveref}

\makeatletter{}
\g@addto@macro\bfseries{\boldmath}
\makeatother

\usepackage[normalem]{ulem}

\newcommand{\be}{\begin{equation}}
\newcommand{\ee}{\end{equation}}
\newcommand{\ba}{\begin{aligned}}
\newcommand{\ea}{\end{aligned}}

\def\Z{{\mathbb{Z}}}
\def\R{{\mathbb{R}}}

\begin{document}

% format
\baselineskip=18pt  % a la harvmac
\numberwithin{equation}{section}  % make eq labels (sec.num)
\allowdisplaybreaks  % allow page breaks in displayed eqs

%%%%%%%%%%%%%%%%%%%%%%%%%%%%%%%%%%%%%%%%%%%
%%%        TITLE BEGINS HERE
%%%%%%%%%%%%%%%%%%%%%%%%%%%%%%%%%%%%%%%%%%%

\thispagestyle{empty}

\vspace*{-2cm}

\vspace*{2.6cm}
\begin{center}
{\Large{\textbf{$(-1)$-Form Symmetries and Anomaly Shifting from SymTFT \vspace{.4cm} \\
}}} 
\vspace*{1.5cm}

Daniel Robbins and Subham Roy\\

{\it Department of Physics, \\
University at Albany, Albany, USA}

\vspace*{0.8cm}
\end{center}
\vspace*{.5cm}

\noindent
We investigate $(-1)$-form symmetries using the framework of symmetry topological field theories. Previous studies of $(-1)$-form symmetries have primarily focused in SymTFTs with topological point operators.  Here we examine SymTFTs devoid of point operators, constructed to realize zero-form symmetries of some physical theory.  In this context we identify codimension-one defects within the bulk of SymTFT constructed via higher gauging which can be interpreted as the generators of $(-1)$-form symmetry of the absolute theory. In addition, we present examples where $(-1)$-form symmetries exhibit the novel ability to shift the 't Hooft anomalies of the theory.

\newpage
%%%%%%%%%%%%%%%%%%%%%%%%%%%%%%%%%%%%%%%%%%%
%%%           TITLE ENDS HERE
%%%%%%%%%%%%%%%%%%%%%%%%%%%%%%%%%%%%%%%%%%%
\setcounter{tocdepth}{2}
\tableofcontents
%\printindex

\newpage

%%%%%%%%%%%%%%%%%%%%%%%%%%%%%%%%%%%%%%%%%%%
%%%        MAIN TEXT BEGINS HERE
%%%%%%%%%%%%%%%%%%%%%%%%%%%%%%%%%%%%%%%%%%%

\section{Introduction}
Symmetries underpin many fundamental principles in Quantum field theories, from organizing the spectrum of operators in gauge theories to constraining renormalization group flows through anomalies. Due to the pioneering work \cite{Gaiotto:2014kfa}, the concept of symmetry has been reformulated in terms of topological operators. Since then, the notion of symmetries has been generalized in various directions -- including higher form symmetries, higher group symmetries and non-invertible symmetries \cite[and references therein]{Freed:2022iao,Cordova:2022ruw,Schafer-Nameki:2023jdn,Bhardwaj:2023kri,Shao:2023gho,Gomes:2023ahz,Brennan:2023mmt,McGreevy:2022oyu,Carqueville:2023jhb,Chang:2018iay,Bhardwaj:2022yxj,Iqbal:2024pee,Costa:2024wks}. 

More recently, the advent of Symmetry Topological Field Theory (SymTFT) \cite{Freed:2012bs,Freed:2022qnc,Gaiotto:2020iye,Kaidi:2022cpf,Kaidi:2023maf,Apruzzi:2021nmk,Bhardwaj:2023ayw} has provided a powerful framework to explore generalized global symmetries. The topological operators of the SymTFT encode the information of symmetry generators and defects charged under these symmetries. Remarkably, the SymTFT allows one to `\textit{isolate}' the notion of symmetry from the physical theory itself, enabling the independent study of these topological operators. 

The notion of $p$-form symmetry, $0 \leq p \leq d-2$ is well explored from the SymTFT and absolute theory. Beyond this `conventional' range, there exists two extreme cases of higher form symmetry corresponding to $(-1)$-form symmetry and $(d-1)$-form symmetry. The $(d-1)$ symmetries are primarily linked with the phenomenon of \textit{decomposition}\cite{Hellerman:2006zs, Sharpe:2014tca, Sharpe:2019ddn, Tanizaki:2019rbk, Komargodski:2020mxz, Cherman:2020cvw, Robbins:2020msp, Nguyen:2021naa, Sharpe:2021srf, Cherman:2021nox, Sharpe:2022ene,Lin:2022xod, Bhardwaj:2023bbf}, $(-1)$-form symmetries -- the lesser known cousin of higher form symmetries -- have recently garnered significant attention \cite{Cordova:2019jnf, Cordova:2019uob,Aloni:2024jpb,Lin:2025oml,Yu:2024jtk,Najjar:2024vmm,Najjar:2025htp, Heidenreich:2020pkc, Vandermeulen:2022edk}. 

From the perspective of the absolute theory, a $(-1)$-form symmetries are naturally associated with space-time filling defects -- operators that would `act' on the theory itself modifying its defining parameters. These ideas have recently been explored within the framework of SymTFT \cite{Lin:2025oml,Yu:2024jtk,Najjar:2024vmm,Najjar:2025htp}. In the SymTFT framework, $(-1)$-form symmetries are associated with codimension-one topological defects that live in the bulk of SymTFT with their natural ability to act on \textit{`dual'} topological point operators of the SymTFT bulk. The topological point operators are associated with the $(d-1)$-form symmetry. When kept parallel to the symmetry boundary (depending on the boundary condition), the codimension one topological defect modifies the values of the point operators on the symmetry boundary of the SymTFT.  As a consequence, we see a change in one of the defining parameters of the absolute quantum field theory. 

In this note, building upon our previous work\cite{Lin:2025oml}, we further investigate $(-1)$-form symmetries through SymTFT. Specifically, in this work, rather than directly constructing a SymTFT to encode $(-1)$-form symmetries and their dual $(d-1)$-form symmetries, we consider SymTFTs without topological point operators. In the absence of point operators, there is no natural dual codimension-one defect within the SymTFT. We will see in detail that in such circumstances, the codimesion-one defect associated with a $(-1)$-form symmetry can arise in a more subtle manner, such as via higher gauging\cite{Roumpedakis:2022aik,Lin:2022xod,Kong_2020,kong2021anyoncondensationtensorcategories,Kong:2014qka,Else_2017,Gaiotto:2019xmp}.  To our knowledge, such a generalization has not been explored before.

In our previous work\cite{Lin:2025oml}, we came up with a generalization of $(-1)$-form symmetries to the non-invertible cases, showing that a $(-1)$-form symmetry is capable of modifying the global symmetries of the theory it acts on. In this note, in addition to the generalization of $(-1)$-form symmetries to SymTFTs without point operators, we also give an example where the $(-1)$-form symmetry modifies the 't Hooft anomaly of the theory it acts on.

The paper is organized as follows. We start with a general discussion of $(-1)$-form symmetries and decomposition, how they are intertwined with each other. Section \ref{sec:section_3} presents a discussion of $(-1)$-form symmetries in the framework of SymTFT. We present the generalization of $(-1)$-form symmetry to SymTFTs without topological point operators. Then in section \ref{sec:section_4}, we construct $2d$ theories with $(-1)$-form symmetries that can shift the 't Hooft anomaly of the theory upon its action. We give further examples of $(-1)$-form symmetries in section \ref{sec:section_5}. Finally, we conclude with future directions and outlook in section \ref{sec:section_6}.

\section{Generalities: $(-1)$-form symmetry and decomposition}\label{sec:section_2}
There are two different approaches to the notion of $(-1)$-form symmetry. One approach is based on decomposition\cite{Lin:2025oml,Yu:2024jtk,Najjar:2024vmm}, and the other is based on the ideas of parameter space\cite{Yu:2024jtk,Cordova:2019jnf,Cordova:2019uob,Aloni:2024jpb}. In this section, we would like to bridge the two different points of view and argue that these two seemingly different perspectives are in coherence with each other.

\subsubsection*{Decomposition Perspective:} 
Decomposition refers to the phenomenon where, in the presence of a $(d-1)$-form symmetry our QFT breaks up into a bunch of \textit{universes}\cite{Hellerman:2006zs, Sharpe:2014tca, Sharpe:2019ddn, Tanizaki:2019rbk, Komargodski:2020mxz, Cherman:2020cvw, Robbins:2020msp, Nguyen:2021naa, Sharpe:2021srf, Cherman:2021nox, Sharpe:2022ene,Lin:2022xod, Bhardwaj:2023bbf}. The partition function of our theory reads as the direct sum of the partition function of individual universes.
\begin{equation}
    \mathcal{T} = \bigoplus_i T_i, \qquad Z_{\mathcal{T}} = Z_{T_1} + Z_{T_2} + ...+Z_{T_i}
\end{equation}
Here $T_i$ denotes the universes. Gauging the $(d-1)$ form symmetry undoes the decomposition\cite{Sharpe:2022ene}, effectively projecting us onto different universes. Each universe carries a $(-1)$-form symmetry, which is the quantum dual of the $(d-1)$-form symmetry and manifests itself as a space-time filling defect. From this perspective, the action of $(-1)$-form symmetry is well understood: it acts on the theory itself and takes us from one universe to another.  

Things become somewhat mysterious when we speak of a theory possessing $(-1)$-form symmetry away from the context of decomposition. For an example, consider the $4d$ Yang-Mills theory.
This is a famous example of $U(1)^{(-1)}$ symmetry obtained by promoting the $\theta$-angle to a space-time dependent field and coupling it to a Chern-Weil-type current \cite{Heidenreich:2020pkc,Aloni:2024jpb}. This is equivalent to treating $\theta(x)$ as the background gauge field of a $(-1)$-form symmetry. Once $\theta$ becomes space-time dependent, we are naturally led to consider a continuous family of Yang-Mills theories, each labeled by a different value of $\theta$. This is where decomposition comes into play; each copy of Yang-Mills labeled by different value of $\theta$ is a universe of a decomposing theory with a $(d-1)$-form symmetry. To uncover the $(d-1)$-form symmetry we need to gauge the $(-1)$-form symmetry, since they are quantum dual of one another. Unlike ordinary symmetries, gauging $(-1)$-form symmetry requires us to take a direct sum of theories related by its action. 

We can make this more concrete. The existence of $(-1)$-form symmetry implies that we are dealing with a finite or infinite collection of theories that are related by its action. We can define a new theory, 
\begin{align}
    \mathcal{T}_D = \bigoplus_i {T}_i
\end{align}
here we have taken a formal direct sum of the individual theories (${T}_i$) that related by the action of $(-1)$-form symmetry. The resulting theory $\mathcal{T}_D$ carries a $(d-1)$-form symmetry. This is a general feature of theories built by taking a direct sums of different theories.

For instance, starting from a given QFT, $X$,  we could now consider a different theory is given by two copies of $X$, which normally we can write as $X \oplus X$.  This is the easiest way to construct a theory with a $(d-1)$-form symmetry, because it has multiple topological local operators, which are the projection operators onto the two theories.  These operators have a natural fusion rule, and this is a generic feature of direct sum theories -- the operators in question do not have any natural group-like structure. To be explicit, let the projection operators be $v_1$ and $v_2$. Then they fuse as,
\begin{equation}
\begin{aligned}
v_1 \otimes v_1 &= v_1, \\
v_2 \otimes v_2 &= v_2, \\
v_1 \otimes v_2 &= v_2 \otimes v_1 = 0
\end{aligned}
\end{equation}
If we wanted we could recombine these operators to give this $(d-1)$-form symmetry a $\Z_2$ group structure, which would require taking $(v_1+v_2)/2$ and $(v_1-v_2)/2$ as our new operators, but doing so is sort of artificial\footnote{We thank Thomas Vandermeulen for discussing this point with us.}.

The very definition of $\mathcal{T}_D$ is equivalent to \textit{gauging $(-1)$-form symmetry}\footnote{We can choose not to sum over all the theories within this space. Such a partial sum corresponds to gauging a subgroup of the $(-1)$-form symmetry.}. In the case of Yang-Mills theory, we have to consider an infinite sum, since the number of universes--labelled by different $\theta$-angle--is infinite. By promoting the $\theta$ angle to a space-time dependent field, we implicitly introduced a collection of Yang-Mills theories. In this setting, the notion of $(-1)$-form symmetry is transparent. Furthermore, upon gauging the full $(-1)$-form symmetry or a subgroup of it we can observe the emergence of a 3-form symmetry\cite{Tanizaki:2019rbk,Yu:2024jtk,Lin:2025oml,Najjar:2025htp}. In a nutshell, whenever there is a $(-1)$-form symmetry, decomposition always lurks in the shadows.

\subsubsection*{Parameter space perspective:} 

There is another approach to understanding $(-1)$-form symmetry, one that does not rely on the concept of decomposition. This scheme, building on the perspective of~\cite{Gaiotto:2020iye} has recently been discussed in \cite{Heckman:2024oot,Santilli:2024dyz}.
From this point of view, a topological manipulation/operation that changes the values of defining parameters of a theory should be regarded as a $(-1)$-form symmetry. Consider a family of QFT's fibered over the parameter space $X$ of the theory, parametrized by $\theta$. While ordinary and higher form symmetries act fiber-wise, the $(-1)$-form symmetries act on the base. A topological operation that takes us from one point of the base to another should be thought of as $(-1)$-form symmetry. We can quantify this as follows, 
\begin{align}
    \text{QFT}_\theta \rightarrow \mathcal{U}_{\theta,\theta'} \text{QFT}_{\theta'}
\end{align}
where $\text{QFT}_\theta$ is QFT defined at $\theta \in X$ and $\mathcal{U}_{\theta,\theta'}$ is the topological operator associated with $(-1)$-form symmetry -- this defect is space-time filling(wraps the QFT direction). The action of this operator is to modify the parameters ($\theta$) of the theory.

Once again, the simplest example is the 4d Yang Mills theory with $\theta$-term. We can imagine the base space as a circle parametrized by $\theta$, with 4d YM theories fibered over it. The $(-1)$-form symmetry then acts by shifting us along the circle, effectively changing the value of $\theta$. 
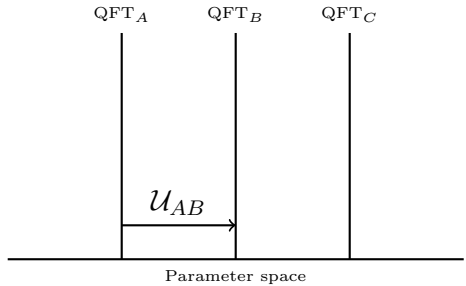
\begin{figure}[H]
    \centering
    \begin{tikzpicture}[scale=1.5]
    % Draw the horizontal line
    \draw[thick] (0,0) -- (4,0) node[midway, below] {\tiny{Parameter space}};
    
    % Draw vertical lines
    \foreach \x/\name in {1/\tiny{$\text{QFT}_A$}, 2/\tiny{$\text{QFT}_B$}, 3/\tiny{$\text{QFT}_C$}} {
        \draw[thick] (\x,0) -- (\x,2) node[above] {\name};
    }
    
    % Fill the space between A and B with light color
    % \fill[blue!20] (1,0) rectangle (2,2);
    
    % Draw arrow from A to B
    \draw[thick, ->] (1,0.3) -- (2,0.3) node[midway, above] {$\mathcal{U}_{AB}$};
\end{tikzpicture}
    \caption{\footnotesize{Imagine the horizontal line as the parameter space. Each vertical line is QFT. For example, these are Pure Yang Mills theories at different $\theta$-value. If we start with QFT A, then by the action of operator $\mathcal{U}_{AB}$ we can move from A to B.} }
    \label{Operator Description}
\end{figure}
In the absence of a continuous parameter like $\theta$, the moduli space of the theory consists of discrete collection of points; equivalently, we can just call it a collection of theories\footnote{A more precise description would be to refer this as a category of QFTs, where a $(-1)$-form symmetry is simply a morphism ($\text{Hom}(\text{QFT}_1, \text{QFT}_2)$) in this category. This concept can be generalized to (-n)-form symmetries by allowing different `types' of morphisms. See \cite[section 2.6]{Yu:2020twi}.}. In this framework, a $(-1)$-form symmetry takes us from one distinct theory in this collection to another. Although there was no mention of decomposition in this approach, we can bring it into the picture by taking a formal direct sum of different theories defined at different points in the parameter space. This is equivalent to gauging the $(-1)$-form symmetry, which naturally results in a $(d-1)$-form symmetry.

\subsubsection*{When can we say that we have $(-1)$-form symmetry?}
The arguments given above tend to suggest that the $(-1)$-form symmetry is not an intrinsic property of the theory; rather it is a reflection of the fact that our theory is a universe of a decomposing theory. Then the action of $(-1)$-form symmetry taking us from one theory to a different theory holds nicely. So,  we can conclude, whenever our theory has a $(-1)$-form symmetry or there exists an operation that relates it with another theory; then our theory must be a universe of a decomposing parent theory carrying a $(d-1)$-form symmetry. This will become even more transparent, when we discuss $(-1)$-form symmetries through the lens of SymTFT in section \eqref{sec:section_3}.

\section{$(-1)$-form symmetry via SymTFT}\label{sec:section_3}
The interplay between $(-1)$-form symmetry and decomposition becomes even more transparent from the perspective of the powerful SymTFT. We will see that two perspectives we discussed earlier blend nicely in SymTFT. In order to explain decomposition from the symmetry TFT we must allow topological point operators in the bulk. These point operators will come with dual codimension-one defects, which should be identified with the generator of $(-1)$-form symmetry.

In the absence of topological point operators in the bulk there are no intrinsic codimension one topological defects in SymTFT. Nevertheless, we can engineer codimension one topological defects in the SymTFT using the Higher gauging techniques \cite{Lin:2022xod,Roumpedakis:2022aik,Kong_2020,kong2021anyoncondensationtensorcategories,Kong:2014qka,Else_2017,Gaiotto:2019xmp}. To begin with, we will allow point operators in the bulk of SymTFT and study the codimension one defect that comes along with it. Then we will tackle the versions obtained by some higher gauging. 
\subsection{SymTFT with Topological Point Operators}
Consider a $2d$ QFT $\mathcal{T}$ with $\Z_N^{(1)}$ symmetry, the SymTFT for such a theory is given by,
\begin{equation}\label{OperatorsInSYM}
    S = \frac{2\pi}{N} \int \theta \cup \delta c_2
\end{equation}
where both $c_2$ and $\theta$ are $\Z_N$-valued cochains. We are working with $2d$ theories for simplicity, but our arguments can be generalized to any dimension. The topological operators in the bulk are given by,
\begin{equation}\label{18}
\begin{aligned}
     U_\alpha &= e^{i\alpha\int c_2}, \qquad  W_\beta = e^{i\beta \theta}, \qquad \alpha, \beta \in \Z_N
\end{aligned}
\end{equation}
We define the physical boundary of our SymTFT as, 
\begin{align}
    \ket{\mathcal{T}} \propto \sum_{c_2} \text{Z}_\mathcal{T}[c_2] \ket{c_2}
\end{align}
where we have coupled our 2d QFT $\mathcal{T}$ and ${Z}_\mathcal{T}$ is the partition function of $\mathcal{T}$. We have two options for our symmetry boundary, we can either choose Dirichlet boundary condition for $c_2$ and Neumann for $\theta$ or we can do vice versa. Going with the former, the symmetry boundary can be defined as, 
\begin{align}
    \bra{D} \propto \sum_{c'_2} \bra{c'_2}\delta(c'_2 - \hat{C}_2)
\end{align}
where `D' stands for Dirichlet boundary condition. We can reduce the sandwich with this boundary condition to obtain the partition function of the original theory $\mathcal{T}$,
\begin{align}
    \bra{D}\mathcal{T}\rangle = \text{Z}_\mathcal{T} [\hat{C}_2]
\end{align}
where $\hat{C}_2$ is the background field for the $\Z^{(1)}_N$ symmetry. We can change the symmetry boundary condition by a discrete fourier transform,
\begin{align}
    \bra{N} \propto \sum_{c_2} \text{exp}(i\int \theta \cup c_2) \bra{c_2}
\end{align}
if we reduce the sandwich with this boundary condition,  
\begin{equation}
\begin{aligned}\label{Startingattheta}
    \bra{N}\mathcal{T}\rangle &= \sum_{c_2,c'_2} \text{exp}(i\int \theta \cup c'_2) \text{Z}_\mathcal{T}[c_2] \bra{c'_2}c_2\rangle \\
    &= \text{Z}_{\mathcal{T}/{\Z_N}}[\theta]
\end{aligned}
\end{equation}
it leads us to the gauged version of the original theory ${\mathcal{T}/{\Z^{(1)}_N}}$, here $\text{Z}_{\mathcal{X}/{\Z_N}}[\theta]$ is the partition function of one individual universe. Now if we add the codimension 1 defect $U_\alpha$ in the bulk parallel to the symmetry boundary\footnote{In terms of boundary condition, this is equivalent to saying, We are imposing Neumann boundary condition on $c_2$ and $\theta$ is kept fixed on the Symmetry boundary. As a result, the operator $U_\alpha$ remains a non-trivial topological operator. In a similar fashion, if we change the boundary condition, we get a theory with non-trivial topological point operators.}, 
\begin{equation}
\begin{aligned}\label{ShiftingTheta}
    \bra{N}U_\alpha\ket{\mathcal{X}} &= \sum_{c_2,c'_2} \text{exp}(i\int \theta \cup c'_2) e^{i\alpha \int c_2} \text{Z}_\mathcal{T}[c_2]  \bra{c'_2}c_2\rangle \\
    &= \sum e^{i\int (\theta +\alpha) \cup c_2} \text{Z}_\mathcal{T} [c_2] \bra{c'_2}c_2\rangle \\
    &= \text{Z}_{\mathcal{T}/{\Z_N}} [\theta + \alpha]
\end{aligned}
\end{equation}
We observe that the value of $\theta$ gets modified in the presence of this defect, which implies the fact that now we are in a different universe labeled by $\left(\theta + \alpha\right)$. We can interpret \eqref{Startingattheta} and \eqref{ShiftingTheta} as two distinct theories being related by the action of $(-1)$-form symmetry. We can summarize the action of $U_\alpha$ purely in terms of boundary condition as, 
\begin{align}
    \bra{N_{\theta}} \longrightarrow \bra{N_{\theta + \alpha}}, \qquad \alpha \in \Z_N
\end{align}
here $\bra{N_\theta}$ denotes the choice of symmetry boundary condition with a specific value of $\theta$. As we emphasized earlier, codimension one defects in the bulk of the SymTFT that links with the bulk point operators, can modify the symmetry boundary non-trivially, by shifting the value of a defining parameter. This is the most traditional version of $(-1)$-form symmetry. In a more realistic setting, when our QFT has other symmetries, our SymTFT will be more involved, there will be additional couplings due to the presence of other symmetries. Nevertheless, the defect $U_\alpha$ will always act on the point operators (in the language of generalized charges\cite{Bhardwaj:2023wzd}, it will carry 0-charges), or equivalently it links with the $W_\beta$ non-trivially, because of that, this type of defect can always modify the value of the $\theta$-angle. 
\subsubsection*{Gauging $(-1)$-form symmetry:} As we have already argued, gauging $(-1)$-form symmetry is equivalent to taking direct sum of theories related by its action. From the SymTFT perspective it is like taking a direct sum of different theories that share the same physical boundary but their symmetry boundary is different. Starting from $\bra{N_\theta}$ as our symmetry boundary, we can generate a plethora of boundary conditions related by the action of $U_\alpha$. Now, if we take a direct sum of all such boundary conditions and define, 
\begin{equation}
    \bra{N^{Sym}_D} = \bra{N_\theta} \oplus \bra{N_{\theta + \alpha}} \oplus \bra{N_{\theta + 2\alpha}} \oplus .....
\end{equation}
reducing the SymTFT with this non-simple boundary condition, 
\begin{equation}
\begin{aligned}
    \bra{N^{Sym}_D} B_{Phys} \rangle &= [\bra{N^{Sym}_\theta} \oplus \bra{N^{Sym}_{\theta + \alpha}} \oplus \bra{N^{Sym}_{\theta + 2\alpha}} \oplus .....] B_{Phys} \rangle] \\
    &= \bra{N_\theta} B_{Phys}\rangle \oplus \bra{N_{\theta + \alpha}} B_{Phys} \rangle \oplus \bra{N_{\theta + 2\alpha}} B_{Phys} \rangle \oplus ..... \\
\mathcal{Z}_D    &= \mathcal{Z}_\theta \oplus \mathcal{Z}_{\theta +\alpha} \oplus \mathcal{Z}_{\theta + 2\alpha } \oplus ......
\end{aligned}
\end{equation}
we end up with a theory $\mathfrak{T}_D$ with a partition function $\mathcal{Z}_D$. We have essentially gauged the $(-1)$-form symmetry, this gauging produces a direct sum of theories, implying Decomposition. In other words, the theory $\mathfrak{T}_D$ has a $(d-1)$-form symmetry. We could have obtained this directly from SymTFT by imposing the Neumann boundary condition on $\theta$, making the point operators in \eqref{OperatorsInSYM} topological.
\subsection{SymTFT without Topological Point Operators: Gauge Defects}
In the previous section, we allowed topological point operators (TPOs) in the bulk of SymTFT and observed how the codimension one defect acts on these point operators. In this section, we turn our attention to discuss SymTFTs that do not contain any TPOs to begin with. This rules out couplings of the form $b_0dc_d$ in the SymTFT action. In a situation like this, the codimension one defects have to show up in a more obscure manner. Specifically, we have to resort to higher gauging techniques \cite{Roumpedakis:2022aik} to identify these defects. Defects constructed by the higher gauging methods -- commonly known as \textit{Condensation defects}\cite{Roumpedakis:2022aik,Lin:2022xod,Kong_2020,kong2021anyoncondensationtensorcategories,Kong:2014qka,Else_2017,Gaiotto:2019xmp} are known to act on extended objects instead of point operators. In doing so, they can also alter the symmetry boundary in a nontrivial way, resulting in a $(-1)$-form symmetry for the absolute theory. 

As an example, we would like to discuss how the codimension one condensation defects of $3d$ $\Z_2$ gauge theory
act on the boundary conditions. This theory can be viewed as the SymTFT corresponding to a $2d$ QFT ($\mathfrak{T}$) with (non-anomalous) $\Z_2^{(0)}$ symmetry. Depending on the choice of symmetry boundary condition,  the resulting absolute theory is either $\mathfrak{T}$ with $\Z_2^{(0)}$-form symmetry, or its orbifold $\mathfrak{T}/\Z_2$, with a quantum dual $\hat{\Z}_2^{(0)}$ symmetry.
\begin{equation}
\begin{aligned}\label{withoutbulkinsertion}
    \bra{B_{Sym}} &= \bra{B_e} \rightarrow \bra{B_e} B_{Phys} \rangle = \mathfrak{T}, \qquad
    \bra{B_{Sym}} =\bra{B_m}  \rightarrow  \bra{B_m} B_{Phys}\rangle = \mathfrak{T}/\Z_2
\end{aligned}
\end{equation}
This change in boundary condition is mediated by a codimension one defect $S_\psi$ \cite{Roumpedakis:2022aik,Kaidi:2022cpf,Gaiotto:2020iye} \footnote{We have reviewed the surface defects of $3d ~\Z_2$ gauge theory in Appendix~\ref{app:A}.}. One can simply insert the defect $S_\psi$ in the bulk and then close the sandwich to change the boundary condition.
\begin{equation}
    \begin{aligned}
        \bra{B_e} S_\psi \ket{B_{Phys}} &= \bra{B_m} B_{Phys} \rangle = \mathfrak{T}/\Z_2, \\
        \bra{B_m} S_\psi \ket{B_{Phys}} &= \bra{B_e} B_{Phys} \rangle = \mathfrak{T}
    \end{aligned}    
\end{equation}

\subsubsection*{Boundary condition changing defect of $\Z_2$ SymTFT:}
We would like to review the construction of this boundary condition-changing defect of this theory explicitly, following \cite{Kaidi:2022cpf}. The action of the $3d$ $\Z_2$ gauge theory is give by:
\begin{equation}
    S_{3d} = {\pi} \int_{M_3} a_1 \delta b_1
\end{equation}
where $a_1,b_1$ are $\Z_2$ valued fields. The topological line operators in the theory are: 
\begin{align*}
    L_{(1,0)}[\gamma_1] = \exp \left(i\pi \int_{\gamma_1} a_1\right), \qquad L_{(0,1)}[\delta_1] = \exp \left(i\pi \int_{\delta_1} b_1\right)
\end{align*}
They generate the electric and magnetic $\Z_2^{(1)}$ symmetry of the 3d SymTFT. We are using the nomenclature used in \cite{Kaidi:2022cpf}, in this notation $L_{(1,0)} = e, ~L_{(0,1)}=m$, where $e$ and $m$ are the electric and magnetic anyons of the $\Z_2$ gauge theory.

The codimension-one defect that can modify the boundary condition from Dirichlet to Neumann can also be interpreted as the defect that interchanges the two gauge fields of the theory, 
\begin{equation}
    a_1 \rightarrow b_1, \qquad b_1 \rightarrow a_1
\end{equation}
Following \cite[Fig.~10]{Kaidi:2022cpf}, we can claim that this defect can be constructed by condensing the line operator $L_{(1,-1)}(\gamma_1) = L_{(1,0)}(\gamma_1) \otimes L_{(0,-1)}(\gamma_1)$;
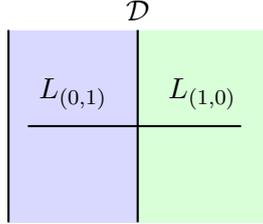
\begin{figure}[H]
    \centering
    \begin{tikzpicture}[scale=0.85]
    
    % Draw lines and fill colors
    \fill[blue!15] (0,0) rectangle (2,3); % Blue region
    \fill[green!15] (2,0) rectangle (4,3); % Green region
    
    % Draw interface (thick black line)
    \draw[thick, black] (2,0) -- (2,3);
    
    % Draw boundary lines
    \draw[thick] (0,0) -- (0,3);
    \draw[thick] (4,0) -- (4,3);
    
    \node at (2.0,3.3) {$\mathcal{D}$}; % Label for interface

    \draw[thick, black] (0.3,1.5) -- (2,1.5);
    \draw[thick, black] (2,1.5) -- (3.6,1.5);

    \node at (3, 2) {$L_{(1,0)}$};
    \node at (1, 2) {$L_{(0,1)}$};
    \end{tikzpicture}
    \caption{\footnotesize{We have an insertion of $\mathcal{D}$ within the $\Z_2$ gauge theory. If we pierce a $e$ line through the defect, it gets modified into a $m$ line.}}
    \label{PierceSpsi}
\end{figure}
when an operator $L_{(1,0)}$ intersects the defect, it can be absorbed by one of the  $L_{(1,-1)}$ lines in the condensate, provided that an $L_{(0,1)}$ line is also emitted, with the net effect that the $L_{(1,0)}$ line becomes an $L_{(0,1)}$ line as it crosses the defect.
\begin{equation}
\label{eq:EMSwapDefect}
    \mathcal{D} = \sum_{\gamma_1 \in H_1(M_2, \Z_2)} L_{(1,-1)}(\gamma_1)
\end{equation}
It is easy to check that $\mathcal{D}$ is indeed invertible.

Certainly, this is not an action of $(-1)$-form symmetry per se\footnote{We can frame all such discrete gaugings as $(-1)$-form symmetries, if we are willing to take a direct sum of the theories connected by discrete gauging. This direct sum procedure introduces decomposition and gives rise to the notion of $(-1)$-form symmetry as the quantum dual symmetry of the $(d-1)$-form symmetry. Of course, this would be a fairly artificial example of $(-1)$-form symmetries; more interesting are cases where $(-1)$-form symmetries arise naturally in some theory under consideration.}, but there exist certain theories in which discrete gauging corresponds to modifying some parameter of the theory, such as the radius of the free boson theory in $2d$ or the complexified coupling constant in $4d$ Maxwell theory\cite{Hasan:2024aow}. In these specific examples, such gaugings can be interpreted as $(-1)$-form symmetry because they correspond to modification of a defining parameter of the theory. 

The $2d$ free boson theory at radius $R$ possesses a $U(1)^{(0)} \times U(1)^{(0)}$ symmetry, with a mixed anomaly between them. We are allowed to gauge a discrete subgroup of the momentum $U(1)^{(0)}$ symmetry, which results in a rescaling of the radius of the free boson theory. This discrete gauging operation is equivalent to an action of $(-1)$-form symmetry. Furthermore, every discrete gauging operation in SymTFT corresponds to a topological interface within the bulk of SymTFT\cite{Kaidi:2022cpf,Gaiotto:2020iye}. So, we can argue that there should be a codimension one defect within the free boson SymTFT that corresponds to this rescaling of the radius $R$, which we identify as the generator of the $(-1)$-form symmetry. This notion of $(-1)$-form symmetry has been discussed previously from the  perspective of $2d$ absolute theory, goes by the name of \textit{gauge defects}\cite{Vandermeulen:2023smx}. In the next subsection, we give a SymTFT perspective of the gauge defects. 
\subsubsection*{Codimension one defects of Free Boson SymTFT:}
The SymTFT for a $2d$ free boson theory with a $U(1)^{(0)} \times U(1)^{(0)}$ symmetry\footnote{The $(d+1)$-dimensional SymTFT for d-dimensional QFT with $U(1)$ global symmetry was constructed using $\R$ gauge fields \cite{Antinucci:2024zjp,Brennan:2024fgj}.
\begin{align}
    S_{\R} = \frac{1}{2\pi} \int b_{d-p} \wedge da_p
\end{align}
Here, $a_p$ and $b_{d-p}$ are $\R$ valued gauge field. The line operators in the bulk are, $U_\alpha[\gamma] = e^{i\alpha\int b_{d-p}}, ~  W_\beta[\gamma] = e^{i\beta\int a_p}, \quad  \alpha ,\beta \in \R$.
We obtain a theory with $U(1)^{(0)} \times U(1)^{(0)}$ symmetry by terminating the defects $U_m$ and $W_n$ with $m,n \in \Z$ on the symmetry boundary. We have specialized to $d=2, p=1$ to get a $U(1)^{(0)} \times U(1)^{(0)}$ symmetry.} is the following \cite{Antinucci:2024zjp,Brennan:2024fgj}:
\begin{equation}\label{SymTFTFB}
\begin{aligned}
    S_{FB} = \frac{1}{2\pi} \int b_1 \wedge da_1
\end{aligned}
\end{equation}
here $a_1$ and $b_1$ are $\R$ valued gauge fields. The operators in the bulk are, 
\begin{equation}
    U_\alpha[\gamma_1] = e^{i\alpha\int b_{1}},  \quad
    W_\beta[\gamma_1] = e^{i\beta\int a_1}, \quad \alpha,\beta \in \mathbb{R}
\end{equation}
As usual, there are two canonical boundary conditions of this SymTFT, where we either terminate all $W_\beta$ or all $U_\alpha$ lines on the symmetry boundary. In both cases, this would give rise to a theory with $\R^{(0)}$ symmetry. As the Pontraygin dual of $\R$ is $\R$, these two boundary conditions are related with each other by a flat gauging. %\footnote{This change of boundary condition can be associated with an interface in the bulk, 
%\begin{equation}
%    \begin{aligned}
%        I_{G} = \exp\left( i \int a_1 \wedge b_1 \right)
%    \end{aligned}
%\end{equation}
%Colliding this interface with the chosen symmetry boundary will result in a flat gauging of the $\R^{(0)}$ symmetry, we end up with a theory with quantum dual $\R^{(0)}$ symmetry.}.
Instead, if we choose to terminate $U_n,W_m, ~\text{with} ~ n,m \in \Z$ on the symmetry boundary, we realize $U(1)^{(0)} \times U(1)^{(0)}$ symmetry in $2d$. These can be identified with the momentum and winding $U(1)$ symmetries of the free boson. The operator $U_\alpha$, which links naturally with $W_m$, will generate the momentum $U(1)_m$. Similarly, $W_\beta$, which links naturally with $U_n$, will generate $U(1)_w$. Moreover, these two $U(1)$ symmetries will have a mixed anomaly, which can be seen from the nontrivial braiding between the generators.  Furthermore, we can choose to terminate $U_{2n}~ \text{and} ~W_{\frac{m}{2}}$ on the symmetry boundary. This is allowed choice of boundary condition as they do give rise to a consistent Lagrangian algebra, due to the trivial braiding among themselves. This choice of boundary condition corresponds to gauging a $\Z_2$ subgroup of the $U(1)$.

There is an alternate formulation of the free boson SymTFT using $U(1)$ and $\R$ gauge fields,
\begin{align}
    S_{FB} = \frac{1}{2\pi} \int b_1 \wedge dB_1 + a_1 \wedge dA_1 - B_1 \wedge dA_1
\end{align}
Here $a_1, b_1$ are $\R$ valued and $A_1,B_1$ are $U(1)$ valued gauge fields. The mixed anomaly between the $U(1)$ symmetries is more transparent in this setting. If we plug in the equation of motion of $A_1$: $da_1 = dB_1$ back in the action, we recover  \eqref{SymTFTFB}. The operators in the bulk are, 
\begin{equation}
    \begin{aligned}
        U^1_\alpha &= \exp \left( i\alpha \oint b_1\right), \qquad W^1_m = \exp \left(im \oint B_1\right) \\
        U^2_\beta &= \exp \left( i\beta \oint a_1\right), \qquad W^2_n = \exp \left(in \oint A_1\right)
    \end{aligned}
\end{equation}
where, $\alpha,\beta \in U(1)$ and $m,n \in \Z$. In this formulation, we need to impose Dirichlet boundary condition on $A_1$ and $B_1$ to realize a theory with $U(1) \times U(1)$ symmetry directly. This is equivalent to saying, $W^1_m, W^2_n$ terminates on the symmetry boundary while $U^1_\alpha , U^2_\beta$ remains parallel. In order to gauge a $\Z_2 \subset U(1)$ the boundary conditions needs to be modified, $W^2_n, W^1_{2m}, U^1_{\frac{p}{2}}$ with $n,m \in \Z$ and $p \in \Z_2$ terminates on the symmetry boundary, while $U^2_\beta, U^1_\alpha, W^1_p$ with $\beta \in \R/\Z \cong U(1)$, $\alpha \in \R/\left(\frac{\Z}{2}\right) \cong U(1)$ and $p\in\Z_2$ remains parallel, under the new boundary conditions. 

In order to change the boundary condition, we need a defect in the SymTFT bulk that will only act on $W^1_1$ and modify it to a $U^1_{\frac{1}{2}}$ line. This operator will be completely inert to everything else. 
\begin{figure}[H]
\centering
    \begin{tikzpicture}[scale=0.85]
    
    %Draw lines and fill colors
    \fill[blue!15] (0,0) rectangle (2,3); % Blue region
    \fill[green!15] (2,0) rectangle (4,3); % Green region
    
    % Draw interface (thick black line)
    \draw[thick, black] (2,0) -- (2,3);
    
    % Draw boundary lines
   \draw[thick] (0,0) -- (0,3);
    \draw[thick] (4,0) -- (4,3);
    
    \node at (2.0,3.3) {\footnotesize{$\mathcal{D}$}}; % Label for interface

    \draw[thick, black] (0.3,1.5) -- (2,1.5);
    \draw[thick, black] (2,1.5) -- (3.6,1.5);

    \node at (3, 2) {{$W^1_1$}};
    \node at (1, 2) {{$U^1_{\frac{1}{2}}$}};
    \end{tikzpicture}
    \caption{\footnotesize{The action of $\mathcal{D}$, modifies a $W^1_1$ line operator into a $U^1_{\frac{1}{2}}$ line.}}%    \label{PierceSpsi}
\end{figure}
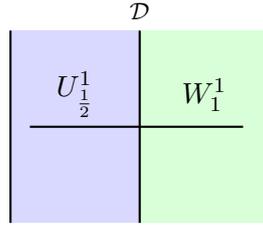
Following the arguments of \cite{Kaidi:2022cpf} that lead to the defect \eqref{eq:EMSwapDefect}, one is tempted to identify $\mathcal{D}$ as the result of condensing the $W^1_1\otimes U^1_{\frac{1}{2}}$ line,
\begin{equation}
    \mathcal{D}(\Sigma_2) \sim \sum_{\gamma_1 \in H_1(\Sigma_2,\Z)} W^1_1(\gamma_1) \otimes U^1_{1/2}(\gamma_1),
\end{equation}
where we are being cavalier about the (possibly infinite) normalization\footnote{This is very analogous to a situation one encounters when attempting to obtain the compact free boson from the non-compact free boson in 2D.  The non-compact boson has an $\R$ translation symmetry, and the compact free boson should be obtained by gauging a $\Z$ subgroup of $\R$.  The usual orbifold procedure for finite groups does not quite work here, since we get factors like $1/|\Z|$ appearing in front of infinite sums, but one can obtain the correct expression for the compact boson partition function if one is willing to ignore these issues in intermediate steps.  Here we are also attempting to gauge an infinite discrete group (again isomorphic to $\Z$), and our hope is that the correct answers can similarly be gleaned by this admittedly non-rigorous argument.  We hope to return to a more careful approach to these issues in future work.}.  In other words we are 1-gauging a $\Z$ subgroup of the $\Z\times U(1)$ 1-form symmetry in the bulk that is generated by the $W^1_m(\gamma)\otimes U^1_\alpha(\gamma)$ lines in the bulk.
Colliding this defect with the symmetry boundary produces the boundary condition required to gauge the $ \Z_2 \subset U(1)_{m} $ of the momentum symmetry in the absolute theory. We can identify this defect with the generator of $(-1)$-form symmetry or gauge defect which modifies the radius of the free boson theory from $R$ to $R/2$.  It would be good to be able to put these arguments, and more general topological manipulations involving infinite discrete groups, on firmer footing.
\subsection{Non-Invertible $(-1)$-form symmetry:}
The generalization of $(-1)$-form symmetries to non-invertible cases have recently appeared in \cite{Lin:2025oml}. While the invertible $(-1)$-form symmetry just shifts a parameter of the theory, its non-invertible counterpart does more. In addition to shifting the parameter of the theory, the action of a non-invertible $(-1)$-form symmetry can turn a simple boundary condition into a non-simple one. Such codimension-one defects with non-invertible fusion rule are also present in the $3d$ $\Z_2$ gauge theory, which can arise as the $3d$ SymTFT of a $2d$ QFT $\mathfrak{T}$ with a $\Z_2^{(0)}$ symmetry. 

\subsubsection*{Non invertible surface defects of 3d $\Z_2$ gauge theory}
In the $3d$ $\Z_2$ gauge theory, there are four non-invertible defects\cite{Roumpedakis:2022aik} with the following fusion rules: 
\begin{equation}
\begin{aligned}
       S_e \otimes S_e &= (\mathcal{Z}_2) S_e, \qquad S_{em} \otimes S_{em} = S_{em},\\  
       S_m \otimes S_m &= (\mathcal{Z}_2) S_m, \qquad  S_{em} \otimes S_{me} = (\mathcal{Z}_2) S_e,
\end{aligned}
\end{equation}
where $\mathcal{Z}_2$ is the partition function of $2d$ $\Z_2$ gauge theory. We briefly review these defects in Appendix \ref{app:A}.  In the language of the canonical boundary conditions of the $3d$ gauge theory,
\begin{equation}
        \begin{aligned}
        \bra{B_e} B_e\rangle &= \bra{B_m} B_m \rangle = (\mathcal{Z}_2), \\
        \bra{B_e} B_m \rangle &= \bra{B_m} B_e\rangle = 1
        \end{aligned}
\end{equation}
These non-invertible defects can be factorized into boundaries as, 
\begin{equation}\label{fusionin3Dgauge}
    S_e = \ket{B_e}\bra{B_e}, \quad S_m = \ket{B_m}\bra{B_m}, \quad S_{em} = \ket{B_e}\bra{B_m}
\end{equation}
Now, we would like to study the action of these defects on the symmetry boundary, we insert the defect in the middle and close the sandwich. Starting with $S_e$, 
\begin{equation}     
        \begin{aligned}
        \bra{B_e} S_e \ket{B_{Phys}} &= \mathcal{Z}_2 \boxtimes \mathfrak{T}, \\
        \bra{B_m} S_e \ket{B_{Phys}} &= \mathfrak{T}
        \end{aligned}
\end{equation}
Repeating the same process with $S_m$ and finally $S_{em}$, 
\begin{equation}
    \begin{aligned}
        \bra{B_e} S_m \ket{B_{Phys}} &= \mathfrak{T}/\Z_2, \qquad
        \bra{B_m} S_m \ket{B_{Phys}} = \mathcal{Z}_2 \boxtimes \mathfrak{T}/\Z_2 \\
        \bra{B_e} S_{em} \ket{B_{Phys}} &= \mathcal{Z}_2 \boxtimes \mathfrak{T}/\Z_2, \qquad
        \bra{B_m} S_{em} \ket{B_{Phys}} = \mathfrak{T}/\Z_2
    \end{aligned}
\end{equation}
We can add more than one copy of the defects in the middle, which gives us two options. Either we can fuse them before we collapse the sandwich or we can drag the collection of defects on top of symmetry boundary one at a time. Regardless, we would get the same result as the following:  
\begin{align}
    \bra{B_e} S_e \otimes S_e \ket{B_{Phys}} &= \bra{B_e} (\mathcal{Z}_2) S_e \ket{B_{Phys}} = \underbrace{(\mathcal{Z}_2)^2 \boxtimes \mathfrak{T}}_{\bar{\mathfrak{T}}}
\end{align}
We observe that each copy of $S_e$ introduces a factor of $(\mathcal{Z}_2)$, inserting $\mathcal{N}$ such defects in the bulk will produce, $(\mathcal{Z}_2)^\mathcal{N} \boxtimes \mathfrak{T}$. Running the same algorithm with $S_m$,
\begin{equation}
    \bra{B_m} S_m \otimes S_m \ket{B_{Phys}} = \bra{B_m} (\mathcal{Z}_2) S_m \ket{B_{Phys}} = \underbrace{(\mathcal{Z}_2)^2 \boxtimes \mathfrak{T}/\Z_2}_{\hat{\mathfrak{T}}}
\end{equation}
inserting $\mathcal{M}$ such defects in the bulk will produce, $(\mathcal{Z}_2)^\mathcal{M} \boxtimes \mathfrak{T}/\Z_2$. On the other hand, inserting multiple copies of $S_{em}$ does not produce any new theory,
\begin{equation}
\begin{aligned}
    \bra{B_e} S_{em} \otimes S_{em} \ket{B_{Phys}} &= \bra{B_e} S_{em} \ket{B_{Phys}} = \mathcal{Z}_2 \boxtimes \mathfrak{T}/\Z_2 \\
    \bra{B_m} S_{em} \otimes S_{em} \ket{B_{Phys}} &= \bra{B_m} S_{em} \ket{B_{Phys}} =  \mathfrak{T}/\Z_2
\end{aligned}
\end{equation}

These defects are codimension-one in the bulk, and hence in the absolute theory they would naturally be interpreted as (noninvertible) $(-1)$-form symmetries.  Unlike the invertible codimension-one defect $S_\psi$ which takes us from one physical theory to another, these non-invertible $(-1)$-form symmetries may take us from a single universe to a decomposing theory with multiple universes.

\subsubsection*{From simple to non-simple boundaries:}
\begin{figure}[H]
    \centering
    \begin{tikzpicture}[scale=1]
    
    % Draw lines and fill colors
    %\fill[blue!15] (0,0) rectangle (2,3); % Blue region
    %\fill[blue!15] (2,0) rectangle (4,3); % Green region
    
    % Draw interface (thick black line)
    \draw[thick, black] (2,0) -- (2,3);
    
    % Draw boundary lines
    \draw[thick] (0,0) -- (0,3);
    \draw[thick] (4,0) -- (4,3);
    
    % Labels
    %\node at (1.0,1.6) {$\mathfrak{Z}(\mathcal{\Z}_2)$};
    %\node at (3,1.6) {$\mathfrak{Z}(\mathcal{\Z}_2)$};

    % Labels
    \node at (0,3.4) {$\mathcal{B}^{Sym}_{\mathcal{\Z}_2}$}; % Label for left line
    \node at (4.2,3.4) {$\mathcal{B}^{Phys}$}; % Label for right line
    \node at (2.0,3.4) {$\mathcal{S}_e$}; % Label for interface

    \draw[thick, black] (0,2) -- (2,2);
    \draw[thick, black] (0,1) -- (2,1);

    \node at (1, 1.3) {$1$};
    \node at (1, 2.3) {$e$};

     % Add equal sign 
    \node at (5,1.5) {=}; % Equal sign
    
    \draw[thick, black] (6,0) -- (6,3); 
    \draw[thick, black] (8,0) -- (8,3);
    
    \filldraw[red] (2,1) circle (3pt);
    \filldraw[red] (0,1) circle (3pt);
    \filldraw[red] (2,2) circle (3pt);
    \filldraw[red] (0,2) circle (3pt);

    \filldraw[red] (6,1) circle (3pt);
    \filldraw[red] (6,2) circle (3pt);

    \node at (6,-0.4) {$\mathcal{B}^{Sym}_{\mathcal{\Z}_2} \otimes \mathcal{S}_e$}; % Label for left line
    \node at (8,-0.4) {$\mathcal{B}^{Phys}$}; % Label for right line

    \node at (9,1.5) {=}; % Equal sign
    \draw[thick, black] (10,0) -- (10,3); 
    
    \node at (11,1.5) {\footnotesize{$(\mathcal{Z}_2) \boxtimes T$}}; % Label for line T
    \end{tikzpicture}
    \caption{\footnotesize{We have condensed the $e$ line to construct $S_e$, which means the $e$ line can now terminate on the defect.}}
    \label{fig:insert_S_e}
\end{figure}
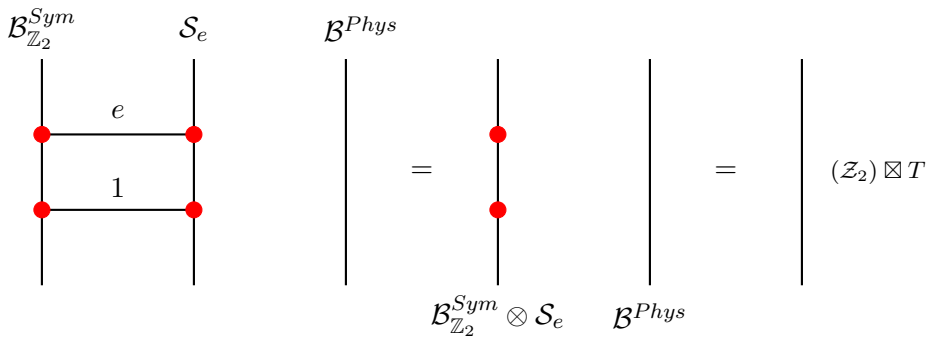

We can see clearly that the action of these non invertible defects on the symmetry boundary is producing nonsimple boundary condition. There is an alternative way to understand this action. The defect $S_e$ was built by condensing the electric anyon of the gauge theory. So, we can say that the electric line can end on the defect. On the other hand, due to the chosen boundary condition, we can imagine the $e$ line to be anchored between the symmetry boundary and the defect, as depicted in Fig \eqref{fig:insert_S_e}. 
When we reduce the sandwich with the defect in the middle, we obtain a set of topological point operators localized on the symmetry boundary. This is an indication that the symmetry boundary is no longer simple. Furthermore, upon reducing to the absolute theory, the presence of these topological point operators indicates decomposition; upon assembling these point operators into projectors, the entire theory decomposes into a direct sum.  We can see that simple boundary conditions become nonsimple in the presence of non-invertible $(-1)$-form symmetries, a feature that has been observed before in \cite{Lin:2025oml}. We can run the same algorithm with $S_m$ or $S_{em}$, insert the defects in the middle and close the sandwich, which will give rise to nonsimple boundary conditions, like before. We summarize the possibilities that give rise to nonsimple boundary conditions in table [\ref{tab:$(-1)$-fstypes}]. 
\begin{table}[H]
            \centering
            \begin{tabular}{|c|c|c|}
                    \hline
                    \textbf{Defect} & \textbf{Initial B.C} & \textbf{Final B.C} \\
                    \hline
                     $S_e$ & $\bra{B_e}$ & $\bra{B_e} \oplus \bra{B_e}$ \\
                    \hline
                    $S_m$ & $\bra{B_m}$ & $\bra{B_m} \oplus \bra{B_m}$ \\
                    \hline
                   $S_{em}$ & $\bra{B_e}$ & $\bra{B_m} \oplus \bra{B_m}$ \\
                    \hline
                    $S_{me}$ & $\bra{B_m}$ & $\bra{B_e} \oplus \bra{B_e}$ \\
                    \hline
            \end{tabular}
            \caption{\footnotesize{Different types of $(-1)$-form symmetry}}
            \label{tab:$(-1)$-fstypes}
        \end{table}

\section{Anomaly shifting $(-1)$-form symmetry}\label{sec:section_4}
In this section, we would like to move our attention to a specific kind of $(-1)$-form symmetry that has the capacity to modify the 't Hooft anomaly across different universes. We first describe this operator in a toy model and then we will give an example of this type of $(-1)$-form symmetry in a $3d$ SymTFT obtained via dimensional reduction of a $5d$ SymTFT. 
\subsubsection*{Toy Model:}
Suppose we have a $2d$ theory $\mathfrak{T}$ with a $\Z_2^{(0)}$ symmetry. The SymTFT of this theory is given by: 
\begin{equation}
    S_{3d} = \frac{i}{\pi}\int a_1 db_1 
\end{equation}
Now consider the theory, $\mathcal{T} = \mathfrak{T} \oplus \mathfrak{T}$. This theory has two topoological point operators, the operators which project onto either of the two copies, and hence $\mathcal{T}$ carries a $\Z_2^{(1)}$ symmetry. We can write down the SymTFT for $\mathcal{T}$ as: 
\begin{equation}
    S^{\mathcal{T}}_{3d} = \frac{i}{\pi} \int a_1db_1 + \frac{1}{\pi}\int b_0 d c_2
\end{equation}
The fields $b_0$ and $c_2$ have been introduced to explicitly accommodate the $\Z_2^{(1)}$.  The two point operators in this language are $1$ and $e^{i b_0}$ (there will be linear combinations that act as projection operators).

The topological operators are:
\begin{equation}
    \begin{aligned}
        X  = e^{i\int b_1}, \quad Y = e^{i\int a_1}, \quad U = e^{ib_0}, \quad V= e^{i\int c_2},
    \end{aligned}
\end{equation}
The $(-1)$-form symmetry is generated by $V$, which moves us between the two copies of $\mathfrak{T}$ by modifying the value of $b_0$ that distinguishes the universes. Now let us repeat this construction for: 
\begin{equation}
    \hat{\mathcal{T}} = \mathfrak{T} \oplus \hat{\mathfrak{T}}
\end{equation}
where $\hat{\mathfrak{T}}$ is a QFT with an anomalous $\Z_2^{(0)}$ symmetry. The SymTFT for $\mathfrak{T}$ and $\hat{\mathfrak{T}}$ are,
\begin{equation}
    \begin{aligned}
    S_\mathfrak{T} &= \frac{i}{\pi}\int a_1db_1, \\ 
    S_{\hat{\mathfrak{T}}} &= \frac{i}{\pi} \int a_1 d b_1 + \frac{1}{2} a_1 d a_1 
    \end{aligned}
\end{equation}
We define the SymTFT for $\hat{\mathcal{T}}$:
\begin{equation}\label{eqn:toy_model_symtft}
    S_{3d} := \frac{i}{\pi} \int a_1 d b_1 + b_0 d c_2 + \frac{1}{2\pi}b_0 a_1 da_1
\end{equation}
We introduce the mixed anomaly to account for the change of 't Hooft anomaly between the universes. This SymTFT must have two distinct boundary conditions, corresponding to the theories $\mathfrak{T}$ and $ \hat{\mathfrak{T}}$. The action is gauge invariant under the following gauge transformations:
\begin{equation}\label{eqn:gauge_t_toy_mod}
    \begin{aligned}
        a_1 &\rightarrow a_1 + d \lambda_0 \\
        b_1 &\rightarrow b_1 + d \mu_0 -na_1 \\
        b_0 &\rightarrow b_0 + 2\pi n \\
        c_2 &\rightarrow c_2 + d \chi_1 - \frac{\lambda_0}{2\pi}d a_1 
    \end{aligned}
\end{equation}
and the equations of motion are: 
\begin{equation}
        \begin{aligned}
        d a_1 &= 0; \qquad d c_2 + \frac{1}{2\pi} a_1d a_1 = 0 \\
        d b_0 &= 0; \qquad
        db_1 +  \frac{1}{2\pi}b_0d a_1 = 0; \\
        \end{aligned}
\end{equation}
The gauge invariant topological operators are:
\begin{equation}\label{eqn:operator_toy_model}
    \begin{aligned}
        X &= \exp\left(i \oint b_1 + \frac{1}{2\pi} b_0a_1\right), \qquad Y = \exp\left(i\oint a_1\right),\\
        U &= \exp \left(i b_0\right), \qquad 
        V = \exp \left(i \int_{\partial M_3} c_2 + \frac{i}{2\pi}  \int_{M_3} a_1 da_1\right)
    \end{aligned}
\end{equation}
The operator $X$ is generator of the $0$-form symmetry that carries some dependence on $b_0$. As a consequence, the self braiding of the line operator depends on the value of $b_0$, in other words the value of self braiding gets modified in each universe. Moreover, we see that the operator $V$, the generator of the $(-1)$-form symmetry becomes non-genuine. This operator can be viewed as an operator that modifies the whole SymTFT action. 

When kept parallel to the symmetry boundary it has two-fold consequences: it modifies the value of $b_0$ through $e^{i\int c_2}$ term and introduces the 3d anomaly inflow term, which adds the anomaly for the 0-form symmetry. The action of the $(-1)$-form symmetry actually modifies the SymTFT of that universe itself. In order to see this explicitly, we define the physical boundary condition of the SymTFT as, 
\begin{align}
    \ket{\hat{\mathcal{T}}} = \sum_{c_2,a_1} Z_{\hat{\mathcal{T}}} [ c_2,a_1] \ket{c_2,a_1}
\end{align}
where $\ket{c_2,a_1}$ is a state in the Hilbert space of the 3d TQFT. We have coupled the theory $\hat{\mathcal{T}}$ to the physical boundary. The symmetry boundary is defined as, 
\begin{align}
    \bra{D} = \sum_{c'_2,a'_1} \bra{c'_2, a'_1} \delta(c'_2 - \hat{C}_2) \delta (a'_1 -\hat{A}_1)
\end{align}
where D stands for Dirichlet boundary condition for the 1-form symmetry and 0-form symmetry backgrounds. Collapsing the sandwich with this boundary condition, gives us the partition function for $\hat{\mathcal{T}}$.
\begin{equation}
\begin{aligned}
    \bra{D} \hat{\mathcal{T}} \rangle &= \sum_{c'_2,a'_1,c_2,a_1}  Z_{\hat{\mathcal{T}}} [ c_2,a_1] \delta(c'_2 - \hat{C}_2) \delta (a_1' -\hat{A}_1) \bra{c'_2} c_2 \rangle \bra{a'_1} a_1 \rangle =Z_{\hat{\mathcal{T}}} [\hat{C}_2,\hat{A}_1]
\end{aligned}
\end{equation}
here $\hat{C}_2$ and $\hat{A}_1$ are the background fields of the 1 form symmetry and the 0 form symmetry. We can gauge the 1-form symmetry by changing the boundary condition via a discrete Fourier transform\cite{Kaidi:2022cpf}, 
\begin{equation}
    \bra{N} = \sum_{c_2',a_1'} \text{exp}\left(\frac{i}{\pi}\int \hat{B}_0 \wedge c_2'\right)\delta(a_1' -\hat{A}_1) \bra{c_2', a_1'}
\end{equation}
where N now stands for Neumann boundary condition (for the one-form symmetry). This change is equivalent to gauging the 1-form symmetry, i.e. undoing decomposition and projecting onto different universes. 
\begin{equation}
    \begin{aligned}
    \bra{N} \hat{\mathcal{T}} \rangle &= \sum_{c_2',a_1',c_2,a_1} \text{exp} \left(\frac{i}{\pi}\int \hat{B}_0 \wedge c'_2\right) Z_{\hat{\mathcal{T}}} [c_2,a_1] \delta(a_1' -\hat{A}_1) \bra{c'_2}c_2\rangle \langle a_1'|a_1\rangle \\
    &= Z_{\hat{\mathcal{T}}/\Z^{(1)}_2} [\hat{B}_0,\hat{A}_1] \\ 
    &= Z_{\mathfrak{T}}[\hat{B}_0, \hat{A}_1]
    \end{aligned}
\end{equation}
If we set $\hat{B}_0=0$, we can identify this partition function as the partition function of the first universe $\mathfrak{T}$, which has a non-anomalous 0-form symmetry(by construction)\footnote{From the SymTFT action it is evident that when $b_0$ is set to $0$, the mixed anomaly vanishes. In fact, the partition function of the absolute theory will have a factor $\exp\left(\frac{i}{2\pi^2}\int b_0 a_1\delta a_1\right)$ outside the path integral, under the gauge transformation of the $a_1$ field, this will produce the anomalous phase but when $b_0$ is set to 0, this term just goes away.}. Now we insert the defect $V$: 
\begin{equation}
    \begin{aligned}
         \bra{N} e^{i\int c_2}e^{\frac{i}{2\pi} \int a_1da_1} \ket{ \hat{\mathcal{T}}}  &= \sum_{c_2} \text{exp} \left(\frac{i}{\pi}\int \hat{B}_0 \wedge c'_2\right) e^{i\int c_2} e^{\frac{i}{2\pi} \int  a_1da_1}  Z_{\hat{\mathcal{T}}} [c_2,a_1] \delta(a_1 -\hat{A}_1)\bra{c'_2} c_2 \rangle \\
        &= \sum_{c_2} \text{exp} \left(\frac{i}{\pi}\int (\hat{B}_0 + \pi) \wedge c_2 \right)  e^{\frac{i}{2\pi} \int  a_1da_1} \delta(a_1 -\hat{A}_1) Z_{\hat{\mathcal{T}}} [c_2,{a}_1] \\
        &= e^{\frac{i}{2\pi}\int \hat{A}_1d \hat{{A}}_1} Z_{\hat{\mathcal{T}}/\Z^{(1)}_2} [\hat{B}_0 + \pi,\hat{A}_1] \\
        &= Z_{\hat{\mathfrak{T}}} [\hat{B}'_0, \hat{A}_1]
    \end{aligned}
\end{equation}
where $\hat{B}_0' = \pi$ as we have set $\hat{B}_0=0$ earlier. In doing so, we can claim this to be the partition function of the second universe $\hat{\mathfrak{T}}$, with an anomalous 0-form symmetry.
\begin{equation}
    \begin{aligned}
        Z_{\hat{\mathfrak{T}}} [\hat{B}_0 = \pi, \hat{A}_1] = e^{i \int \hat{A}_1d \hat{A}_1} \underbrace{Z_{\hat{\mathcal{T}}/\Z^{(1)}_2} [ \hat{B}_0=\pi,\hat{A}_1]}_{\mathfrak{T}~\text{at}~ b_0= \pi}
    \end{aligned}
\end{equation}
If we perform a gauge transformation of the 0-form symmetry, a phase appears as expected, signaling an anomaly. 

\subsubsection*{Example via Dimension Reduction}
In this section, we demonstrate that the SymTFTs with a structure similar to \eqref{eqn:toy_model_symtft} can arise naturally via dimensional reduction. As an illustration, we present a SymTFT featuring a codimension one defect that generates a $(-1)$-form symmetry and has the capacity to modify the 't Hooft anomaly across universes. The proposed SymTFT takes the form, 
\begin{align}\label{eq:3DDimRedAction}
    S_{Sym}^{3D} = \frac{1}{2\pi}\int_{M_3}\left( b_1dA_1 + f_2dB_0 + f_0 dB_2 + \frac{k_{2}}{4\pi} f_0A_1dA_1\right).
\end{align}
where the lowercase letters are $\R$-valued gauge fields and the uppercase letters denote $U(1)$-valued gauge fields.  This SymTFT can be obtained via $S^2$ reduction of a $5d$ SymTFT of a $4d$ $U(1)$ gauge theory with a $U(1)$ flavor symmetry. Moreover, there is a self anomaly for the flavor symmetry. The $5d$ SymTFT of that captures this is given by, 
\begin{equation}
    S^{5d}_{bulk} = \frac{1}{2\pi} \int_{\mathcal{W}_5} \left(b_3dA^\text{4d}_1 + f^\text{4d}_2dB^\text{4d}_2 + \frac{k_{2}}{4\pi} A^\text{4d}_1dA^\text{4d}_1f^{4d}_2\right).
\end{equation}
A similar action first appeared in \cite{Nardoni:2024sos}, where the couplings responsible for ABJ anomaly and 2-group symmetry were also present\footnote{check appndix \ref{app:B} for a brief review.}. Reducing the five dimensional action on $S^2$ with the following ansatz, 
\begin{align}
    \int_{S^2}b_3=b_1,\qquad\int_{S^2} f^\text{4d}_2 = f_0,\qquad\int_{S^2} B^\text{4d}_2=B_0 \, , \qquad \int_{S^2} dA_1 = 0 \, .
\end{align}
in doing so, we recover the three dimensional SymTFT \eqref{eq:3DDimRedAction}.

We may neglect the term $f_2dB_0$, as it is not coupled with other terms in the action. In doing so, we realize that the action matches precisely with the action in \eqref{eqn:toy_model_symtft}. The gauge transformations of this action, analogous to those in \eqref{eqn:gauge_t_toy_mod}, are given by:
\begin{equation}
    \begin{aligned}
        f_0 &\longrightarrow f_0 , \\
        A_1 &\longrightarrow A_1 + d\lambda_0, \\
        b_1 &\longrightarrow b_1 + d\mu_0, \\
        B_2 &\longleftrightarrow B_2 + d\chi_1 - \frac{k_2}{4\pi}\lambda_0dA_1
    \end{aligned}
\end{equation}
which in turn predict a corresponding set of operators,
\begin{equation}\label{eqn:op_non_gen}
    \begin{aligned}
        X[\gamma_1] &= \exp \left( i\alpha \oint_{\gamma_1} b_1\right), \qquad Y[\delta_1] = \exp \left( in\oint_{\delta_1}A_1\right)\\
        U[x_0] &= \exp \left(i\beta f_0 (x_0)\right), \qquad V[\Sigma_2=\partial M_3, M_3] = \exp \left( i p \int_{\Sigma_2}B_2 + \frac{i p k_2}{4\pi}\int_{M_3} A_1dA_1\right)
    \end{aligned}
\end{equation}
here $\alpha, \beta \in U(1)$ and $p,n \in \Z$. The operators have a similar structure to those found in \eqref{eqn:operator_toy_model}. The operators $U$ and $V[\Sigma_2,M_3]$ may have non-trivial commutation due to the linking between $\Sigma_2$ and $x_0$,
\begin{equation}
    \begin{aligned}
        U[x_0]V[\Sigma_2,M_3] = \exp \left( 2\pi i \beta p \right) V[\Sigma_2,M_3] U[y_0],
    \end{aligned}
\end{equation}
where $x_0$ is a point to the left of the $V[\Sigma_2,M_3]$ and $y_0$ is a generic point to the right of $V$. The above linking relation implies that as we take the point operator $U[x_0]$ through $V$, the value of $f_0$ jumps by $2\pi p$. 

We are going to impose the Dirichlet boundary condition on $f_0$, $A_1$ and the Neumann boundary condition on $b_1$ and $B_2$. Due to the chosen boundary condition, we will end up with a $U(1)^{(0)}$ symmetry. Meanwhile the operator $V[\Sigma_2,M_3]$ will generate the $(-1)$-form symmetry. 

If we are in the universe where $f_0$ is set to 0, we have a $U(1)$ 0-form symmetry generated by $X[\gamma_1]$. As the value of $f_0$ jumps due to the $U/V$ linking, we move to the universe with $f_0 = 2 \pi p$ by the action of $V[\Sigma_2,M_3]$.  In this universe, we still have the $U(1)^{(0)}$ symmetry but now due to the Chern-Simons term, there is a self anomaly for the  0-form symmetry is introduced, which can be detected from the self-braiding of the $X[\gamma_1]$ lines, 
\begin{equation}
    \langle X_\alpha(\gamma_1), X_\beta(\gamma_2) \rangle = \exp \left(-2\pi i \alpha \beta k_2 p ~\text{Link}(\gamma_1,\gamma_2)\right) 
\end{equation}
where we have used the fact that $f_0 = 2\pi p$.

\section{$(-1)$-form symmetry in Orbifold groupoid}\label{sec:section_5}
Examples of $(-1)$-form symmetry naturally arise in the context of the orbifold groupoid\cite{Gaiotto:2020iye}. In broad brush, we can think of orbifold groupoid as a collection of theories obtained by orbifolding a quantum field theory with global symmetry, with different choices of discrete torsion. Each choice of discrete torsion leads to a distinct orbifold theory and these variants are related with each other by some topological manipulation, e.g. discrete gauging. 

Once again, instead of being completely abstract we will work with an example in mind. Let us consider a $2d$ theory $T$ equipped with a non-anomalous 0-form symmetry $G = \Z_p \times \Z_p$, with $p$ prime. We can gauge this symmetry and construct an orbifold theory $[T/G]$. Since $H^2(\Z_p \times \Z_p, U(1)) = \Z_p$, we have the freedom to add discrete torsion and obtain a different variant of the orbifold theory. In doing so, we can construct distinct orbifold theories $[T/G]_{\nu^{i}_2}$ ($\nu_2$ is the choice of discrete torsion) depending on the characterization of the discrete torsion. 

This family -- including the original theory $T$ as well as orbifolds by any one of the $p+1$ different $\Z_p$ subgroups -- forms what is referred as the orbifold groupoid. With the groupoid at our disposal, we are allowed to take direct sum of the theories living in this groupoid, e.g.~
\begin{equation}\label{directsumtheory}
\begin{aligned}
    \mathfrak{T}_D &= [T/G]_{\nu^{0}_2} \oplus [T/G]_{\nu^{1}_2} \oplus [T/G]_{\nu^{2}_2} \oplus .....= \bigoplus_i ~ [T/G]_{\nu^{i}_2}
\end{aligned}
\end{equation}
We observe that each universe in this direct sum is distinguished by a specific choice of discrete torsion parameter. Due to the direct sum, the theory $\mathfrak{T}_D$ inherits a $1$-form symmetry. As a consequence, the theory $\mathfrak{T}_D$ decomposes and in principle every theory/universe in this collection should carry a $(-1)$-form symmetry. 

In order to identify this, let us consider the SymTFT of the theory $T$, with a physical boundary, 
\begin{align}
    \ket{B_{Phys}} = \sum_{a_1} Z_T(a_1) \ket{a_1}
\end{align}
here, $\ket{a_1}$ is a basis in the hilbert space of the 3d TFT, $Z_T$ is the partition function of the theory $T$ and $a_1$ is the background gauge field of $G$. As all the theories within this groupoid are obtained by gauging (with or without discrete torsion), the bulk SymTFT must remain unchanged for all the theories in this family. Moreover we can claim that all the theories within this family share the same physical boundary, only the symmetry boundary distinguishes them from each other. 

In order to obtain the orbifold theories we need to impose Neumann boundary condition. Our symmetry boundary to be,
\begin{align}
    \bra{B^{0}_{Sym}} = \sum_{a_1} e^{i\int a_1 \cup b_1}\bra{a_1}
\end{align}
where $b_1$ is the gauge field of $\hat{G}$, the Poyntragin dual of $G$. An interval compactification at this stage would give us the orbifold theory $[T/G]$. More generally, depending on the characterization we are allowed to stack $G$-SPT phases/Discrete torsion to our symmetry boundary, 
\begin{align}
    \bra{B_{Sym}^{1}}= \sum_{a_1} e^{i\int \nu_2(a_1)} e^{i\int a_1 \cup b_1}\bra{a_1} 
\end{align}
where $\nu_2$ is the SPT phase or we can think of this as adding discrete torsion before gauging the $1$-form symmetry on the symmetry boundary. 
\begin{align}
   \bra{B_{Sym}^{2}}= \sum_{a_1} e^{i\int \nu'_2(a_1)} e^{i\int a_1 \cup b_1}\bra{a_1} 
\end{align}
Reducing the sandwich with different symmetry boundaries will produce different variants of the orbifold theory,
\begin{equation}
\begin{aligned}
    Z_{[T/G]_{\nu_2}} &= \langle B_{Sym}^{1}| B_{Phys} \rangle = \sum_{a_1} e^{i\int \nu_2(a_1)} Z_{[T/G]}[b_1] \\
    Z_{[T/G]_{\nu'_2}} &= \langle B_{Sym}^{2}| B_{Phys} \rangle = \sum_{a_1} e^{i\int \nu'_2(a_1)} Z_{[T/G]}[b_1]
\end{aligned}
\end{equation}
We can write down an expression relating the two such $\hat{G}$ symmetric theories, 
\begin{align}
    Z_{[T/G]_{\nu'_2}} = \underbrace{ \sum_{a_1} e^{ i\int \nu'_2(a_1)}e^{- i\int \nu_2(a_1)}}_{\text{$(-1)$--form symmetry}} Z_{[T/G]_{\nu_2}}
\end{align}
We can already see that there exists a topological manipulation that successfully `connects' two distinct theories living within the groupoid. Hence, if we do define the direct sum as we have done in \eqref{directsumtheory}, we can interpret this topological manipulation as $(-1)$-form symmetry\footnote{If we define our universes as $T$ and $[T/G]$ then, the $(-1)$-form symmetry will be analogous to gauge defects.}. In the same spirit, we can take various direct sums using the theories connected within this orbifold groupoid. These will be related with each other via discrete gauging and one can then frame all such discrete gauging operations (morphisms) that connect different nodes of the groupoid on the same footing as $(-1)$-form symmetries.

The theories living within this groupoid are well-defined quantum field theories and they have their own SymTFTs. On the other hand, $\mathfrak{T}_D$ defined by taking direct sum of all these theories will also have its own SymTFT. At the level of action, this SymTFT must carry $\frac{ip}{2\pi} \int_{M_3} b_0 \wedge dc_2$ coupling, which will ensure the existence of topological point operators in the bulk, $\mathcal{U} = e^{i\alpha  b_0}$ along with codimension one defect $\mathcal{V} = e^{i\alpha \int c_2}$. We can project down to specific orbifold theory by imposing Neumann boundary condition on $c_2$. For completeness,
\begin{align*}
    b_0 \mapsto &\text{Neumann}, ~ c_2 \mapsto \text{Dirichlet} \rightarrow \mathfrak{T}_D \\
    b_0 \mapsto &\text{Dirichlet}, ~ c_2 \mapsto \text{Neumann} + \text{Discrete Torsion} \rightarrow \text{universes}
\end{align*} 
\section{Conclusions}\label{sec:section_6}
In this note, we extended the notion of $(-1)$-form symmetries to the framework of SymTFTs in the absence point operators. We connect our construction of $(-1)$-form symmetries with gauge defects\cite{Vandermeulen:2023smx}- a construction that is primarily understood within the context of absolute theories. In the absence of point operators, we argued that the codimension one defects can still arise as condensation defects using the higher-gauging procedure. As a consequence, these do not carry $0$-charges but they do carry $1$-charge (they act on line operators); in doing so, they can modify the symmetry boundary condition nontrivially. We emphasize that while our construction is physically motivated, a more rigorous treatment -- particularly with respect to the normalizations -- remains an important direction for future work.

As a concrete example, we investigated the boundary condition changing defects in the context of free boson SymTFT. It would be particularly interesting to explore analogous codimension one defects in higher dimensional theories, such as four-dimensional Maxwell theory. 

Moreover, we have also demonstrated that certain incarnations of $(-1)$-form symmetries can modify topological data such as the 't Hooft anomalies of the absolute theory. Through the lens of decomposition, $(-1)$-form symmetry can change the 't Hooft anomaly between different universes. These `anomaly shifting' versions of $(-1)$-form symmetry also appear in 3d ABJ(M) theories arising from M2 brane probing Calabi-Yau singularities\cite{vanBeest:2022fss,Yu:2024jtk}\footnote{In addition to the change of 't Hooft anomaly, the $(-1)$-form symmetry in this case modifies the gauge rank of the theory.}. We look forward to addressing these questions in future work. 

\subsection*{Acknowledgments}
We are grateful to Ling Lin, Thomas Vandermeulen for carefully reading the draft. We also thank Shani Meynet and Danielle Miglioratti for helpful discussions. 
\appendix
\section{Surface defects of 3d $\Z_2$ Gauge Theory}\label{app:A}
We would like to review the properties of the codimension one defects of 3d $\Z_2$ gauge theory in this appendix. The action of the 3d $\Z_2$ gauge theory is given by, 
\begin{equation}
    S_{3d} = \frac{i}{\pi} \int_{M_3} a_1 db_1
\end{equation}
where $a_1,b_1$ are $\Z_2$ valued fields. This theory carries a $\Z^{(1)}_{2e} \times \Z_{2m}^{(1)}$ 1-form symmetry. This theory also possesses extended defects, which are constructed via 1-gauging the 1 form symmetry. There are 6 different surface defects associated with six different gauging's of the 1 form symmetry. 
\begin{table}[H]
            \centering
            \begin{tabular}{|c|c|c|}
                    \hline
                    \textbf{Subgroup of} $\Z_2^{(e)} \times \Z_2^{(m)}$ &   \textbf{Defect} & \textbf{Type} \\
                    \hline
                    Trivial Subgroup &  id & Invertible\\
                    \hline
                    $\Z_2^{(e)}$ & $S_e$ & Non-invertible \\
                    \hline
                    $\Z_2^{(m)}$ & $S_m$ & Non-invertible \\  
                    \hline
                    $\Z_2^{(e)} \times \Z_2^{(m)}$ & $S_{em}$ & Non-invertible \\
                    \hline
                    ${(\Z_2^{(e)} \times \Z_2^{(m)})}_{d.t}$ & $S_{me}$ & Non-invertible\\
                    \hline
                    Fermionic Subgroup & $S_\psi$ & Invertible \\
                    \hline
            \end{tabular}
            \caption{6 possible gauging's and the associated Defect. In the fifth row, d.t stands for Discrete Torsion.}
            \label{tab:GaugingsofZ2}
        \end{table}
As we have listed in Table \eqref{tab:GaugingsofZ2}, among invertible defects we have the trivial defect and the boundary condition-changing defect, aka \textit{electro-magnetic duality} defect. The rest are non-invertible defects. We would like to go over some of the key features of these defects. In \cite{Roumpedakis:2022aik}, it was shown that the defect $S_\psi$, with $S_\psi \otimes S_\psi = 1$, has many interesting properties. One of them is its ability to change boundary conditions.
\begin{equation}
\begin{aligned}
    S_\psi \otimes \ket{B_e} &= \ket{B_m}, \qquad S_\psi \otimes \ket{B_m} = \ket{B_e}
\end{aligned}
\end{equation}
where $\ket{B_e}$ and $\ket{B_m}$ are the canonical boundary conditions of the 3d gauge theory obtained by condensing by the electric or magnetic anyon. The operator has the ability to change the boundary condition under fusion.
Another key feature that will be relevant for us is how this defect acts on the anyons\footnote{There are 4 lines in the $\Z_2$ gauge theory. We are using the notation of $\{ 1, e, m, em\}$}. The details were already worked out in \cite{Roumpedakis:2022aik}, we list the results:
\begin{equation}
    \begin{aligned}
        S_\psi \cdot 1= 1, \quad S_\psi \cdot e = m , \quad S_\psi \cdot m = e , \quad S_\psi \cdot em = em
    \end{aligned}
\end{equation}
here $\cdot$ denotes the \textit{piercing action} of this defect. This translates to the fact that if we take the electric $e$ line through the surface it is modified to the $m$ line. 

Apart from $S_\psi$ (and the identity), the rest of the defects are non-invertible and are obtained by 1-gauging different subgroups of the 1-form symmetry. For example, the defect $S_e$($S_m$) is constructed by condensing the line operator $L_{(1,0)}$($L_{(0,1)}$) over a surface\footnote{We are using the notation of \cite{Kaidi:2022cpf}, where $L_{(1,0)}$ is the electric anyon, more traditionally labeled as $e$.}. 
\begin{equation}
    S_e(\Sigma_2) = \sum_{\gamma_1 \in H_1(\Sigma_2,\Z_2)} L_{(1,0)}(\gamma_1)
\end{equation}
We report the relevant fusion rules of these defects in the following, from \cite[See section 6.3 for a full list]{Roumpedakis:2022aik}: 
\begin{equation}
    \begin{aligned}
        S_e \times S_e &= (\mathcal{Z}_2)S_e,  \\
        S_m \times S_m &= (\mathcal{Z}_2)S_m,  \\
        S_e \times S_m &= S_{em}, \\
        S_{em} \times S_{me} &= (\mathcal{Z}_2) S_{e}
    \end{aligned}
\end{equation}

\section{Dimension Reduction of $5$d SymTFT}\label{app:B}
The 5d SymTFT for a 4d $U(1)$ gauge theory with a $U(1)^{(0)}_F$ flavor symmetry is given by,
\begin{align}
    S_\text{bulk}^\text{5d} = \frac{1}{2\pi} \int_{{\cal W}_5} \left(b_3dA^\text{4d}_1 + f^\text{4d}_2dB^\text{4d}_2 + \frac{k_{A}}{4\pi}A^\text{4d}_1(f^\text{4d}_2)^2 + \frac{k_{2}}{4\pi} A^\text{4d}_1dA^\text{4d}_1f^{4d}_2 + \frac{k_T}{12\pi} A^\text{4d}_1(dA^\text{4d}_1)^2\right).
\end{align}
Unlike the action discussed in the main text, here we also added the couplings for the ABJ anomaly and the 2-group symmetry. This action appeared in \cite{Nardoni:2024sos}. Here, $b_3$ and $f_2$ are $\R$-valued, while $A_1$ and $B_2$ are $U(1)$-valued gauge fields, with the following gauge transformations:
\begin{align}\label{eq:gauge-transformations-5dSYMTFT}
\begin{split}
    A^\text{4d}_1\ \longrightarrow\ & A^\text{4d}_1+d\lambda_0\, ,\\
    B^\text{4d}_2\ \longrightarrow\ & B^\text{4d}_2+d\lambda_1+\frac{k_A}{2\pi}A^\text{4d}_1\mu_1-\frac{k_{A}}{2\pi}f^\text{4d}_2\lambda_0+\frac{k_{A}}{2\pi}d\lambda_0\mu_1 \, ,\\
    f^\text{4d}_2\ \longrightarrow\ & f^\text{4d}_2+d\mu_1 \, ,\\
    b_3\ \longrightarrow\ & b_3+d\mu_2-\frac{k_{A}}{2\pi}f^\text{4d}_2\mu_1-\frac{k_{2}}{4\pi}dA^\text{4d}_1\mu_1-\frac{k_{2}}{4\pi}f^\text{4d}_2d\lambda_0-\frac{k_T}{12\pi}dA^\text{4d}_1d\lambda_0-\frac{k_{A}}{4\pi}\mu_1d\mu_1\, .
\end{split}
\end{align}
The $U(1)$ gauge symmetry is reflected having Dirichlet/Neumann boundary conditions for $B_2^\text{4d} / f_2$ on the symmetry boundary \cite{Antinucci:2024zjp}, and the $U(1)_F^{(0)}$ flavor symmetry has background field $A_1^\text{4d}$ which has Dirichlet boundary conditions.

Reducing the theory on $S^2$ with the reduction ansatz,
\begin{align}
    \int_{S^2}b_3=b_1,\qquad\int_{S^2} f^\text{4d}_2 = f_0,\qquad\int_{S^2} B^\text{4d}_2=B_0 \, , \qquad \int_{S^2} dA_1 = 0 \, .
\end{align}
where the authors set $f_0$, which takes values $2\pi m_G \in 2\pi \mathbb{Z}$. Which leads us to the $3d$ bulk action, 
\begin{align}
    S_{Sym}^{3d} = \frac{1}{2\pi}\int_{M_3}\left( b_1dA_1 + f_2dB_0 + f_0 dB_2 + \frac{k_A}{4\pi} f_0 A_1 f_2 + \frac{k_{2}}{4\pi} f_0A_1dA_1\right).
\end{align}
Setting $k_A = 0$ above leads us to the action we started with in \eqref{eq:3DDimRedAction}.

\providecommand{\href}[2]{#2}\begingroup\raggedright\endgroup

%\bibliographystyle{JHEP}
%\bibliography{refs}
\end{document}